\documentclass[letterpaper,twocolumn,prl,superscriptaddress,notitlepage]{revtex4-1}
\usepackage[latin9]{inputenc}
\usepackage{amsmath}
\usepackage{graphicx}
\usepackage{xcolor}
\begin{document}

\title{Quasiparticle spectra from molecules to bulk}

\author{Vojt\v{e}ch Vl\v{c}ek}
\email{vojtech@chem.ucla.edu}

\affiliation{Department of Chemistry and Biochemistry, University of California,
Los Angeles California 90095, U.S.A.}

\author{Eran Rabani}
\email{eran.rabani@berkeley.edu}

\affiliation{Department of Chemistry, University of California and Materials Science
Division, Lawrence Berkeley National Laboratory, Berkeley, California
94720, USA}

\affiliation{The Raymond and Beverly Sackler Center for Computational Molecular
and Materials Science, Tel Aviv University, Tel Aviv, Israel 69978}

\author{Daniel Neuhauser}
\email{dxn@ucla.edu}

\affiliation{Department of Chemistry and Biochemistry, University of California,
Los Angeles California 90095, U.S.A.}

\date{\today}
\begin{abstract}
A stochastic cumulant GW method is presented, allowing us to map the evolution of photoemission spectra, quasiparticle
energies, lifetimes and emergence of collective excitations from molecules
to bulk-like systems with up to thousands of valence electrons, including
Si nanocrystals and nanoplatelet. The quasiparticle energies rise due to their
coupling with collective shake-up (plasmon) excitations, and this
coupling leads to significant spectral weight loss (up to 50\% for
the low energy states), shortening the lifetimes and shifting the spectral features to lower energy by as much as 0.6~eV. Such
features are common to all the systems studied irrespective of their
size and shape. For small and low dimensional systems the surface plasmon resonances affect the frequency of the collective excitation and position of the satellites.
\end{abstract}
\maketitle

Recent developments in Green's function (GF) techniques have allowed for the description of charge excitations, i.e., quasiparticles (QPs) \cite{FetterWalecka,Gross1991}, in the bulk, over a wide range of QP energies. Band-edge excitations are well-described by the so called $G_0W_0$ approximation \cite{Hedin1965,HybertsenLouie,martin2016interacting}, while at higher QP energies corrections are required to account for charge-density fluctuations and hole-plasmon coupling \cite{Aryasetiawan1996,Guzzo2011valence,kas2014cumulant,martin2016interacting}. Photoemission experiments on solids reveal significant QP lifetime shortening and coupling  to other collective excitations, manifested by satellite structures in the photoemission spectra \cite{langreth1970singularities,Aryasetiawan1996,lischner2015satellite}. The  satellite structure and the QP lifetime shortening is often captured by the cumulant expansion (CE) ansatz to $G_0W_0$ \cite{Aryasetiawan1996,Guzzo2011valence,kas2014cumulant,lischner2015satellite,caruso2015band,vigil2016dispersion,McClain2016,Mayers2016} .

In confined systems, the QP spectrum near the band-edge is governed by the quantum confinement of electrons and holes. Higher energy, satellite-excitations are attributed to simultaneous ionization and excitation of the valence electrons (``shake-up" excitations) \cite{Schirmer1983,Cederbaum1986,Deleuze1997,McClain2016, Krylov2017}.  Transition and differences between the satellite spectral features of molecules and nanostructures with ``shake-up" signatures and bulk with collective plasmon resonances have been difficult to assess, as they require many-body treatment of systems with hundreds and thousands of electrons.  In fact, the quantum confinement effect on the satellite transitions has received little attention if any.

In this letter, we address this challenge by combining the well-known cumulant expansion
(CE) ansatz \cite{lundqvist1967single,langreth1970singularities,kas2014cumulant,martin2016interacting,Lischner2013,caruso2015band,Guzzo2011valence}  with the recent stochastic GW approach (sGW \cite{Neuhauser2014,vlcek2016stochastic}), to obtain a 
nearly linear-scaling algorithm that reveals the changes of the QP spectra from a single molecule to covalently bonded nanocrystals (NCs) of unprecedented size (here up to 5288 valence electrons). The formalism is presented and assessed for the two size extremes (molecule and bulk), followed by the study of the effects of quantum confinement on the satellite structure in silicon NCs of different size and shape. In small NCs the satellite features are affected by the changes in the plasmon energy. For large NCs, we find observable quantum confinement effects on the satellite features below the exciton Bohr radius, where the position of the satellite peak and the QP lifetime show small dependence on the size of the system.

The central theoretical quantity for quasiparticles is the spectral
function, which in the sudden approximation is directly linked to
the photoemission current \cite{Aryasetiawan1998,OnidaReiningRubio2002,martin2016interacting}.
The spectral function of the $i^{{\rm th}}$ QP state is $A_{i}\left(\omega\right)=\frac{1}{\pi}{\rm Im}G_{i}\left(\omega\right)$,
where the Green's function fulfills the Dyson equation $G_{i}\left(\omega\right)=G_{i}^{\left(0\right)}\left(\omega\right)+G_{i}^{\left(0\right)}\left(\omega\right)\Sigma_{i}\left(\omega\right)G_{i}^{\left(0\right)}\left(\omega\right)+\cdots$, where
 $G_{i}^{\left(0\right)}\left(\omega\right)$ is the non-interacting Green's function
and $\Sigma_{i}$ is the self-energy. All the quantities are non-local in space and all the higher terms in the equation represent a convolution integral, but for brevity we omit the spatial dependence
in the notation.  

As usual, the non-interacting
system is described by the Kohn-Sham DFT \cite{HohenbergKohn,KohnSham} (see details in \footnote{The LDA DFT calculation used a real-space grid and Troullier-Martins
pseudopotentials. The QP energies were calculated by the $sGW$ approach
detailed in \cite{Neuhauser2014,vlcek2016stochastic}. The grid spacings
were 0.6$a_{0}$ for the Si nanocrystals, 0.4$a_{0}$ for the PH$_{3}$
and NH$_{3}$ molecules and 0.5$a_{0}$ for the C$_{2}$H$_{2}$
molecule. The self-energy is approximated to be diagonal in the basis
of KS eigenstates.}). 
The self energy is then given in the diagonal $G_{0}W_{0}$ approximation
as \cite{Hedin1965}:
$\tilde\Sigma_{i}\left(t\right)=i\left\langle \phi_{i}\middle|\tilde{G}_{i}^{\left(0\right)}\left(t\right)W\left(t^+\right)\middle|\phi_{i}\right\rangle$,
where $t^+$ is infinitesimally after $t$,  $\phi_{i}$ is the KS eigenstate, $W\left(\omega\right)=\epsilon^{-1}\left(\omega\right)v_{c}$,
$v_{c}$ is the Coulomb kernel and $\epsilon^{-1}\left(\omega\right)$
is the inverse dielectric function. 
Quantities in frequency ($G$ and $\Sigma$) and time domains ($\tilde G$ and $\tilde \Sigma$) are simply related by their Fourier transforms.
From the calculated $\Sigma_{i}(\omega)$ the $G_{0}W_{0}$
spectral function is given by:

\begin{equation}
A_{i}^{GW}\left(\omega\right)=\frac{1}{\pi}\frac{\left|{\rm Im}\Sigma_{i}\left(\omega\right)\right|}{\left(\omega-\varepsilon_{i}-{\rm Re}\Sigma_{i}(\omega)+\overline{v}_{XC}\right)^{2}+({\rm Im}\Sigma_{i}\left(\omega\right))^{2}},\label{eq:GW_specfunc}
\end{equation}
where $\varepsilon_{i}$ is the KS eigenstate energy and $\overline{v}_{XC}$
is the expectation value of the mean-field exchange-correlation potential.
$A_{i}^{GW}$ has peaks at the quasi-particle energies, $\varepsilon_{i}^{qp}$,
that fulfill the fixed-point equation, 
\begin{equation}
\varepsilon_{i}^{qp}=\varepsilon_{i}+{\rm Re}\Sigma_{i}\left(\omega=\varepsilon_{i}^{qp}\right)-\overline{v}_{XC}.\label{eq:QPfixed_point}
\end{equation}
In this GW approximation, the inverse lifetime of the QP is given
by the imaginary part of the self-energy at the peak. However, the
actual plasmon-hole coupled excitations are not in general represented
by the isolated poles in Eq.~\eqref{eq:GW_specfunc} and $A^{GW}(\omega)$
thus does not yield a proper description of satellite structures.
In addition, spurious secondary peaks arise if Eq.~\eqref{eq:QPfixed_point}
has multiple solutions \cite{Aryasetiawan1996,Lischner2013,Guzzo2011valence}.

The CE formulation is required to account for the effect of hole-plasmon coupling. For the $i^{{\rm th}}$ occupied state,
the Green's function in the CE \emph{ansatz} reads
\cite{lundqvist1967single,langreth1970singularities,martin2016interacting}:

\begin{equation}
\tilde G_{i}\left(t\right)=-ie^{i\varepsilon_{i}t}e^{C_{i}\left(t\right)}\theta(-t)=-ie^{i\varepsilon_{i}t+C_{i}^{qp}\left(t\right)}e^{C_{i}^{s}\left(t\right)}\theta(-t),\label{Gfull_cumul}
\end{equation}
where $C_{i}$ is the cumulant, obtained from the Dyson series expansion.
Further, following Ref.~\cite{Aryasetiawan1996} the cumulant contribution
is separated into two components. The first is a QP cumulant, $C_{i}^{qp}$,
derived explicitly in Ref. \cite{Aryasetiawan1996} and associated
with a portion of the spectral function describing the main QP peak:

\begin{equation}
A_{i}^{qp}\left(\omega\right)=\frac{{\bf Z}_{i}}{\pi}\frac{\left|{\rm Im}\Sigma\left(\varepsilon_{i}^{qp}\right)\right|}{\left(\omega-\varepsilon_{i}^{qp}\right)^{2}+({\rm Im}\Sigma\left(\varepsilon_{i}^{qp}\right))^{2}},
\end{equation}
where the renormalization factor, due to redistribution of the spectral
weight into the satellite peaks, is ${\bf Z}_{i}=e^{\alpha_{i}}$
with $\alpha_{i}=\left.\frac{\partial\Sigma\left(\omega\right)}{\partial\omega}\right|_{\omega=\varepsilon_{i}}$.
The lifetime of the QP is $1/\left|{\rm Im}\Sigma\left(\varepsilon_{i}^{qp}\right)\right|$. 

\begin{figure}[t]
\includegraphics[width=2.5in]{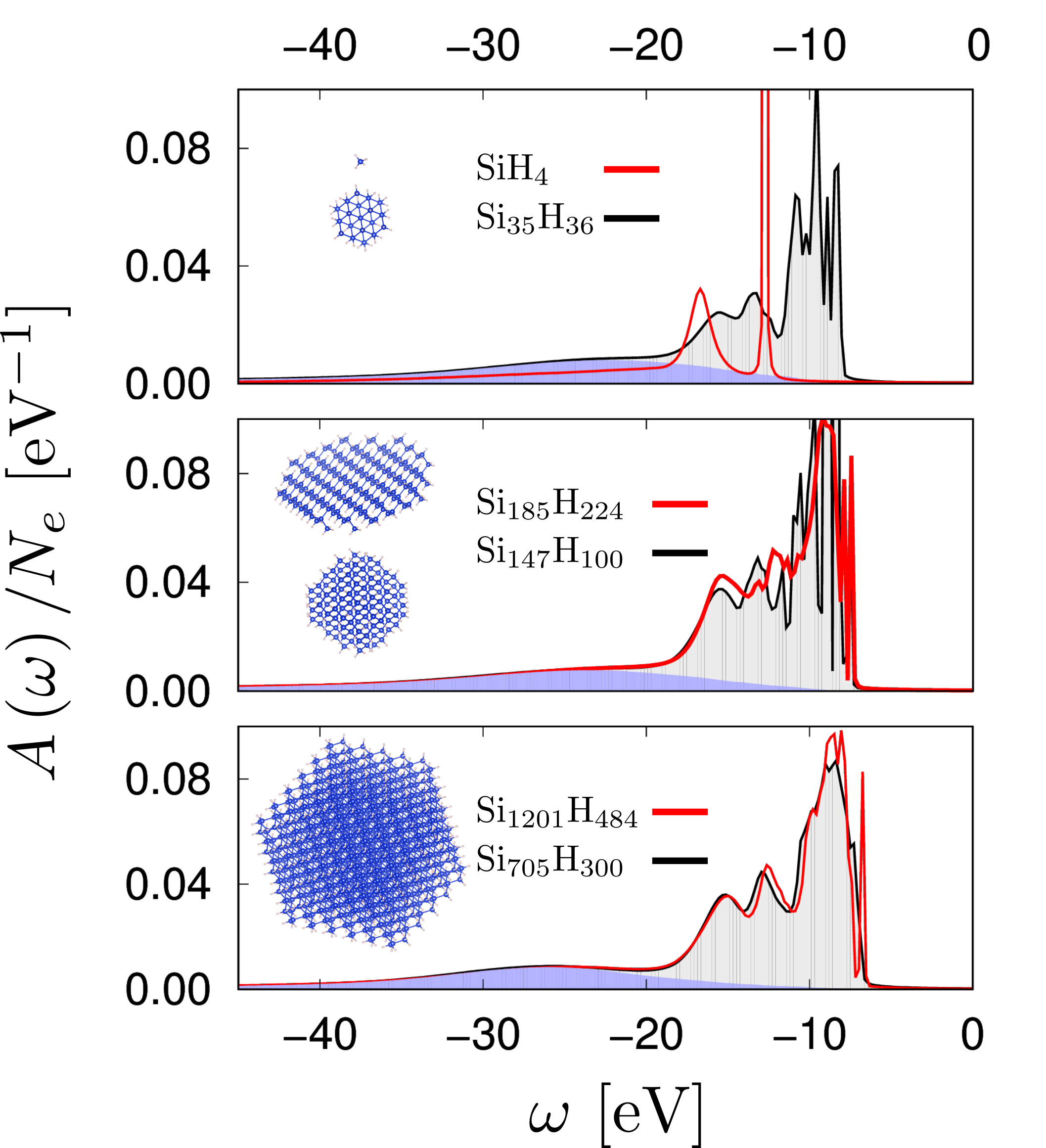} \caption{Spectral functions for silicon nanocrystals, a platelet (middle panel
- red) and a silane molecule. Satellite contributions to the spectral
function are shown by a blue shaded area.}
\label{SiNCs} 
\end{figure}

By itself, $A_{i}^{qp}(\omega)$ does not include any satellite contributions
\textendash{} it is a single Lorentzian-shaped peak around the QP
energy. The satellite peaks stem from resonances identified as poles
in $W\left(\omega\right)$ (i.e., zeros of $\epsilon\left(\omega\right)$) and appear
as strong maxima in the imaginary part of the self-energy; they are
introduced by the second part of the exponential in Eq.~\eqref{Gfull_cumul}
that derives from the spectral representation of $\Sigma_{P}$ \cite{lundqvist1967single,langreth1970singularities,Aryasetiawan1996}:
\begin{equation}
C_{i}^{s}\left(t\right)=\frac{1}{\pi}\lim_{\eta\to0}\int_{-\infty}^{\mu}\frac{{\rm Im}\Sigma_{P}\left(\omega\right)e^{-i\left(\omega-\varepsilon_{i}+i\eta\right)t}}{\left(\omega-\varepsilon_{i}+i\eta\right)^{2}}{\rm d}\omega.\label{eq:Cist}
\end{equation}
We solve for $C_{i}^{s}\left(t\right)$ using $\Sigma_{P}$, obtained
from the stochastic $G_{0}W_{0}$ calculation. 
The computed
satellite cumulant, $C_{i}^{s}\left(t\right)$ , is inserted to Eq.~\eqref{Gfull_cumul}
which is Fourier transformed to yield $G_{i}(\omega)$, and thereby
$A_{i}(\omega)={\rm Im({\it G_{i}{\rm (}\omega{\rm ))}}}$ \footnote{In principle, the total spectral function requires the calculation
of $A_{i}(\omega)$ for all occupied states, but this is numerically
prohibitive except for very small systems. However, since the quasiparticle energies and the spectral functions vary smoothly with frequency [Vlcek et al. 2017, ArXiv:1701.02023] for the extended system studied here, we compute the spectral function for several selected states and interpolate
the result. The number of states is increased till the interpolation
is converged to within 0.1~eV  To evaluate the total spectral function of the large systems in Fig.~\ref{SiNCs},
a single Lorentzian peak was used to describe the satellites. We found
that a third order polynomial fit to the quasiparticle energies and
the parameters of the Lorentzian peaks is sufficient to yield QP energies
and satellite positions within 0.4~eV, i.e. way better than the resolution
of the predicted spectral functions. For systems up to Si$_{705}$H$_{300}$,
we found that calculations for 5 independent states provide converged
results. For Si$_{1201}$H$_{484}$ 3 calculations were performed. }.

\begin{figure}[h]
\includegraphics[width=2.2in]{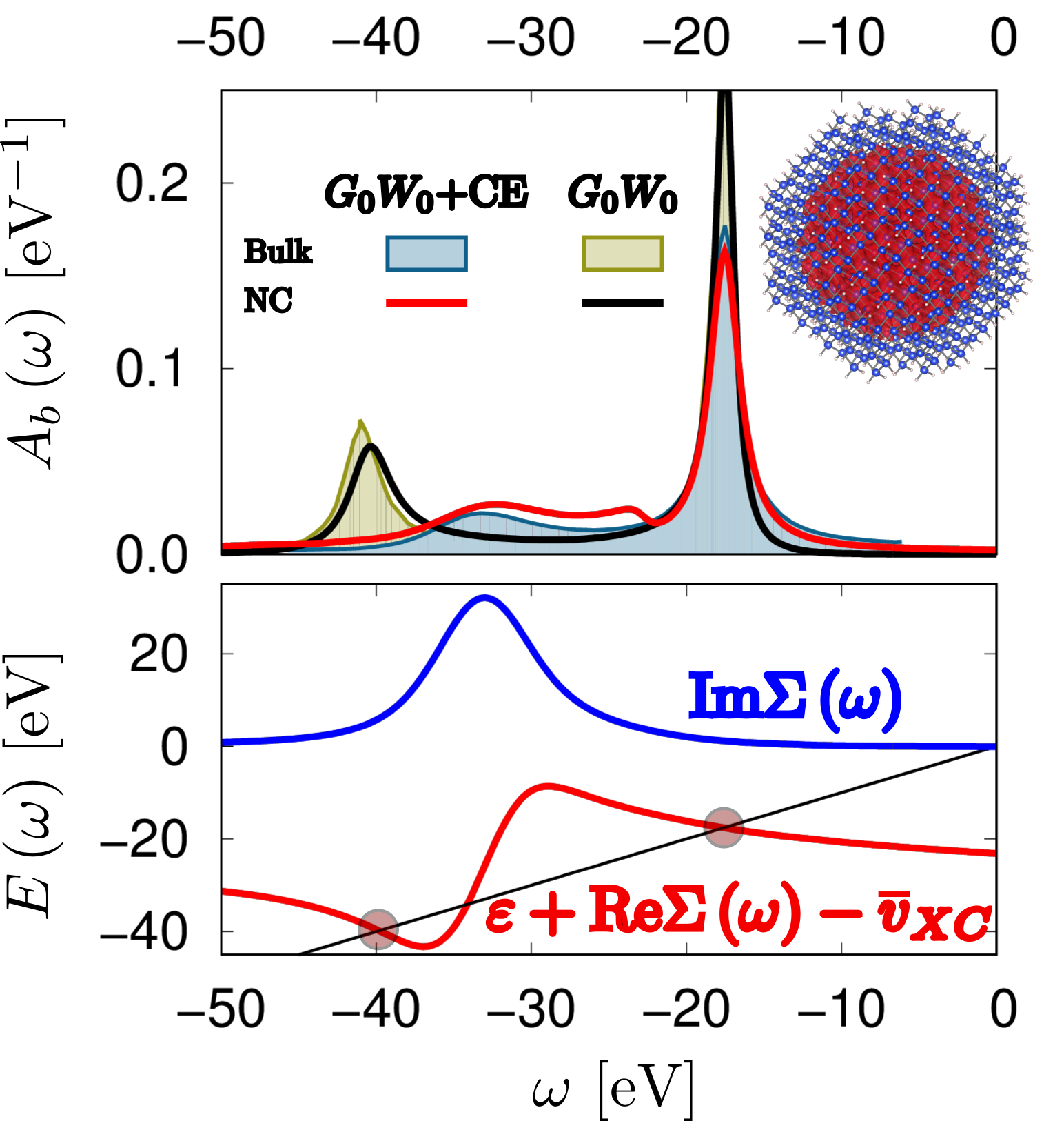} \caption{\textbf{Top}: Spectral function for the bottom VB, $A_b\left(\omega\right)$, for a bulk solid from Ref.~\cite{lischner2015satellite} and for
Si$_{705}$H$_{300}$ NC. The $G_{0}W_{0}$+cumulant spectral
function (red) has an asymmetric satellite peak at the maximum of
${\rm Im}\Sigma(\omega)$. The $G_{0}W_{0}$ prediction (black) has
an artificial second maximum at low energies due to a spurious additional
solution of Eq.~\eqref{eq:QPfixed_point}. The inset plots the structure of the nanocrystal and orbital density isosurface
(red; Si and H are blue and white circles). \textbf{Bottom}: Graphical solutions to the QP equation, marked
with red circles, are found at the intersection of the red curve ($\varepsilon+{\rm Re}\Sigma(\omega)-\bar{v}_{XC}$)
with the diagonal $\omega$ line.}
\label{cumul_effect} 
\end{figure}

We next verify our approach using a large NC, Si$_{705}$H$_{300}$,
that is close to the bulk limit. Fig.~\ref{cumul_effect} shows the
spectral function of the bottom valence band  (VB - denoted $A_{b}^{GW}$) with a pronounced QP
peak at $-17.5$~eV. If a cumulant expansion is not used, $A_{b}^{GW}$
shows an additional maximum at $-39.8$~eV. This is in excellent agreement with previous GW calculations for bulk systems, but is not observed experimentally, and is attributed to spurious secondary solutions to Eq.~\eqref{eq:QPfixed_point} \cite{Aryasetiawan1996,Guzzo2011valence,kas2014cumulant,caruso2015band,lischner2015satellite}. 

With the cumulant $GW$ (Eq.~\eqref{Gfull_cumul}) the spectrum changes drastically
and an additional peak is obtained at $-32.3$~eV in excellent agreement with a result for bulk Si \cite{lischner2015satellite}. This peak is physically
meaningful as it corresponds to the maximum of ${\rm Im}\Sigma(\omega)$
associated with a collective excitation of the valence electrons (plasmon).
The appearance of the satellite structure is accompanied by reduction
of the intensity of the main QP peak, so that the renormalization factor
is ${\bf Z}=0.61$, i.e., 39\% of the intensity is transferred to
the satellites. The asymmetry of the satellite is due to the difference
between the effective masses of the QP and the plasmon \cite{vigil2016dispersion}.
The pronounced transfer of the spectral weight to the plasmon satellite
for the bottom valence excitations is a consequence of their high
energy and spatial extent (leading to large overlaps with other states).
An isosurface of the bottom valence orbital of Si$_{705}$H$_{300}$
indeed exhibits spherical symmetry and lacks nodal planes as seen from the inset of Fig.~\ref{cumul_effect}. 

\begin{figure}[h]
\includegraphics[width=2.5in]{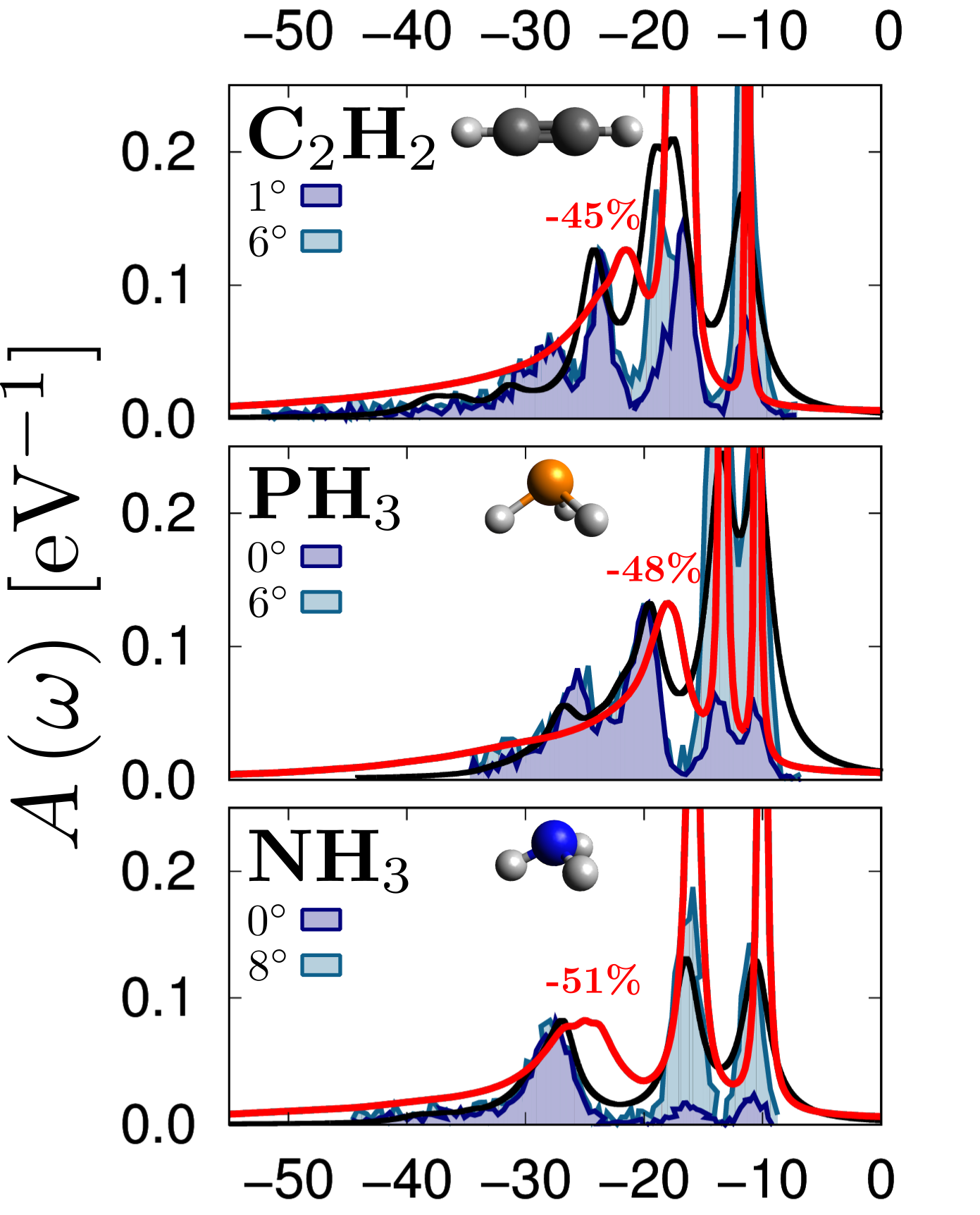} \caption{Spectral functions (red line) from stochastic $G_{0}W_{0}$+CE for C$_{2}$H$_{2}$,
PH$_{3}$ and NH$_{3}$. The spectral weight loss from the bottom
valence state to the satellites is shown above the peak. Theoretical
spectra obtained with SAC-CI \cite{Wasada1989,Ishida2002} are in
black and the colored areas refer to the experimental photoemission
spectra for two relative azimuthal angles \cite{Weigold1991,Ishida2002}. }
\label{molecules} 
\end{figure}

To further test our approach on finite systems, we applied (Fig.~\ref{molecules})
the stochastic $G_{0}W_{0}$ approximation with CE to a series of
small molecules for which experimental photoemission spectra are available.
The results in Fig.~\ref{molecules} were
further scaled so that the bottom valence state peak has the same
intensity as the $G_{0}W_{0}$+CE curve. 
The $G_{0}W_{0}$+CE description relies on the concept of plasmon, valid for extended systems. It is thus surprising that
this approach provides a qualitative description for such small systems. Indeed, the stochastic
$GW$ with damped real-time propagation of the exited state \cite{Neuhauser2014,vlcek2016stochastic}
is in qualitative agreement with experiment and with high level
SAC-CI (symmetry-adapted-cluster configuration interaction) calculations,
computationally feasible for small molecules \footnote{Similar to Ref.~\cite{Ishida2002},
the SAC-CI results were convoluted with a Lorentzian peak with 2~eV
broadening}. We note that:

(i) the QP energies at the top valence band are captured well by $G_0W_0$. This is the energy region where DFT is a good starting point. But $G_0W_0$ fails, however, to reproduce the bottom end of the valence band, where it underestimates the position of the peaks by a significant amount of 2~eV. For these states, DFT is not a good starting point and the
``single-shot" $G_0W_0$ procedure is inaccurate.

(ii) Most importantly, the $G_{0}W_{0}$+CE description captures
the satellite overall decay, although without the fine structure peaks
in the satellite region.  The pronounced satellite spectral weight
comes at the expense of the QP peaks which transfer up to $51\%$
of their intensity to the satellite tails. 
The broadening of the satellite peaks observed in $G_0W_0$+CE is a consequence of an intrinsic decay of the density--density correlation function in time ($\tau$). The peak width is independent of the maximal time used to simulate the screening (we have varied the propagation time from 1 to 24~fs without affecting the lifetime), leading to a set of broad poles in the dielectric function. In theory, an infinite propagation time would result in sharp poles due to recurrences in the correlation function. As clearly can be seen in Fig.\ref{molecules}, accounting only for $\tau$ yields a photoelectron spectrum in good agreement with experiment, likely due to other mechanism suppressing the recurrences in photoelectron spectroscopy.

Further, the $G_{0}W_{0}$+CE spectral function has maxima that are shifted
with respect to the $G_{0}W_{0}$ QP energies. The shift is large
for the bottom valence states; e.g., for NH$_{3}$ the $G_{0}W_{0}$
peak is at $-25.0$~eV while the $G_{0}W_{0}$+CE maximum is at
$-25.7$~eV. The $0.7$~eV difference is significant as it is 17\%
of the GW correction to the LDA energy ($-20.8$~eV). Thus, the usual
practice where $G_{0}W_{0}$ results are directly compared to photoionization
experiment is problematic, especially for low energy states, as it
does not include the coupling of these states to the shake-up excitations. 


In the next, main, part of this letter we investigate the evolution of the spectral function with system size; the results for a series of Si NCs (normalized by the number of electrons) are shown in Fig.~\ref{SiNCs}.  All NCs exhibit a discrete and narrow spectrum near the top of the VB. Due to the quantum confinement effect, the top of the VB shifts to higher energies with increasing size; the highest occupied state has energies of -8.1~eV and -6.4~eV for for Si$_{35}$H$_{36}$ and for Si$_{1201}$H$_{484}$, respectively. For deeper hole excitations, the sharp features merge into a semi-continuous spectral response with significant life-time shortening. This is accompanied by significant spectral weight transfer
($\sim50\%$) to the satellites. 
The bottom of the VB depends weakly on the system size, spanning an energy between -17.3 and -17.7~eV for the range of NCs studied. The QP peak also overlaps with the emerging satellite, which is already well-developed into its bulk shape for Si$_{35}$H$_{36}$ and found in the range typical for bulk silicon \cite{Guzzo2011valence}. This result is rather surprising, since both the QP spectrum near the band edge and the plasmonic excitations are sensitive to the system size.  We further observe that the dimensionality  does not strongly affect the main QP peaks: the silicon platelet has $\sim60\%$ of the Si
atoms on the surface, yet its spectral function is similar to the NCs.

On closer inspection, we observe that the satellite maximum exhibits non-monotonic shifts: First it shows a strong decrease in energy for systems from Si$_{35}$H$_{36}$ to Si$_{705}$H$_{300}$ (from $-22.5$ to $-26.1$~eV, respectively), which is followed by slight move  back to higher energies by 0.6~eV. The initial regime stems from the decrease in the plasmon resonance frequency ($\omega_p$) discussed below in detail. Once $\omega_p$ converges, the satellite maximum follows the changes in the QP DOS of the valence states governed by quantum confinement, i.e.~the spectrum moves to higher energies (c.f., Fig.~\ref{SiNCs}).

\begin{figure}
\includegraphics[width=2.5in]{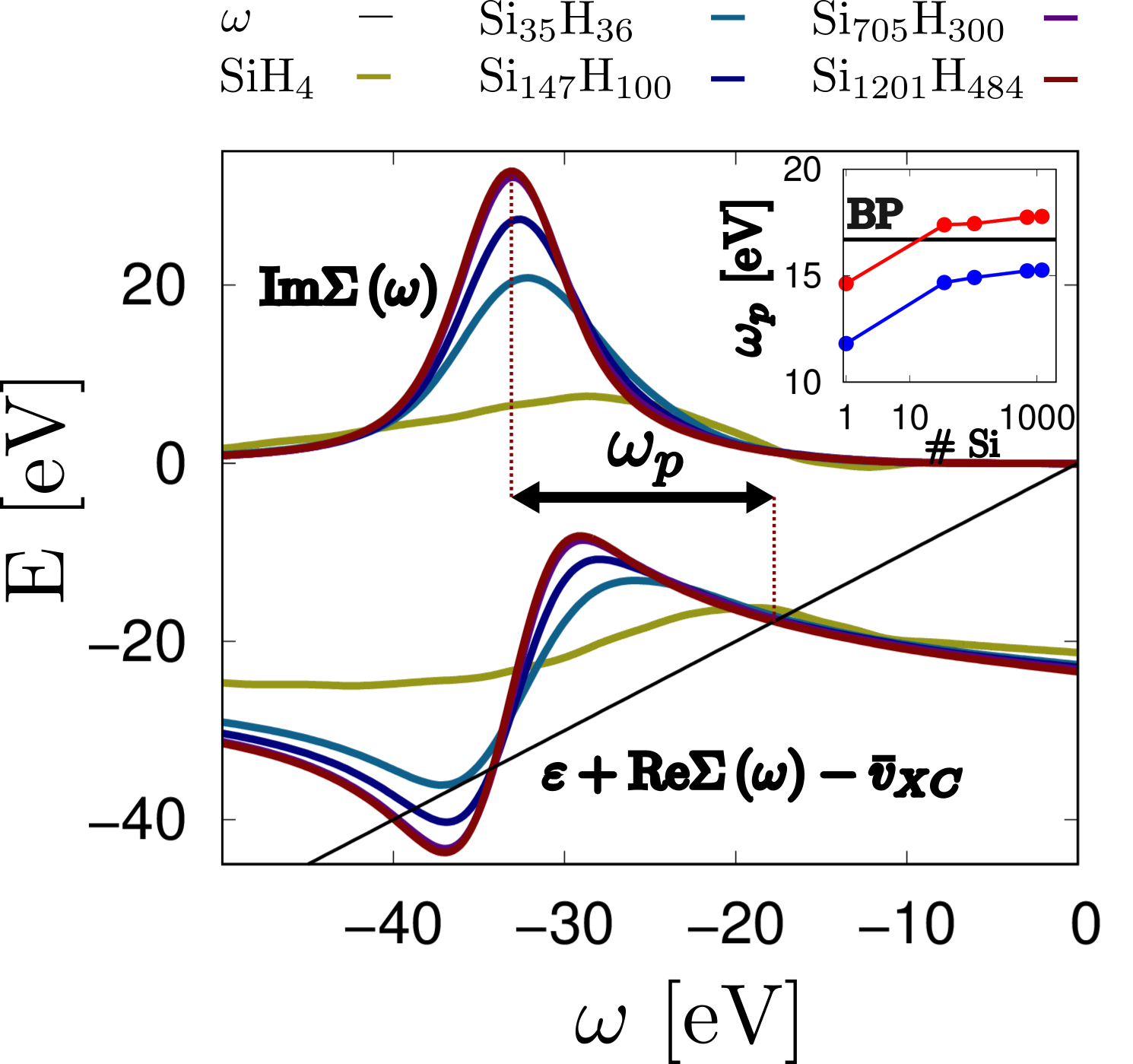}

\caption{Results for the bottom VB of
Si nanocrystals of different sizes: Upper lines (above E=0) show
the imaginary part of the self-energy. Lower curves below E=0 give the shifted real part of the self energy
and represent the graphical solution to Eq.~\ref{eq:QPfixed_point}
(the QP energy is the intersection with the frequency
line in black). The plasmon resonance ($\omega_p$) for the bottom VB hole is shown in the inset for the bottom VB (blue) and for the top VB (red) together with the experimental value of the bulk and surface plasmon (BP - Ref.~\cite{Stiebling1978}) shown by a black vertical line.}\label{fig:Spectral-function-size}
\end{figure}

In Fig.~\ref{fig:Spectral-function-size} we show ${\rm Im}\Sigma_{P}(\omega)$ together with the graphical solution to Eq.~\ref{eq:QPfixed_point} (which also depicts spurious secondary
solutions found already for Si$_{35}$H$_{36}$ at $-36$~eV). The plasmon peak in the ${\rm Im}\Sigma_{P}(\omega)$ curve changes till the asymptotic limit is reached; ultimately the curve for NCs with 3120 electrons (Si$_{705}$H$_{300}$) and 5288 electrons (Si$_{1201}$H$_{484}$) have practically identical height, width and position. The distance between the maximum of ${\rm Im}\Sigma_{P}(\omega)$ and the QP energy corresponds to $\omega_p$ coupled to the bottom valence hole; for the largest system $\omega_p=15.3$~eV. Convergence of $\omega_p$ with system size is shown in the inset for the top and the bottom VBs. 

Unlike in solids, the holes in finite systems couple to two types of plasmon resonances: low energy surface plasmon and high energy bulk plasmon. For small NCs, both contribute and lead to a broad peak in ${\rm Im}\Sigma_{P}(\omega)$. The surface plasmon resonances also strongly contribute in low dimensional structures -- the plasmon satellite of the large Si platelet has a maximum at $-22.6$~eV which is almost identical to the smallest NC (Si$_{35}$H$_{36}$). For big systems, the hole becomes more localized inside the NC (c.f., inset of Fig.~\ref{cumul_effect}) and the coupling to the bulk plasmon dominates, leading to increase in $\omega_p$. The distribution of the resonances becomes more narrow and the peak in ${\rm Im}\Sigma_{P}(\omega)$ decreases in width. 

In summary, our calculations are the first ab-initio theoretical predictions
of the photoemission spectra, quasiparticle energies and lifetimes
covering the wide region between molecules and solids. The calculations
show that the QP energies gradually increase with system
size and this is accompanied by changes in the position of the satellite
peaks which corresponds to a simultaneous ionization of the system
and creation of a collective (shake-up or plasmon) excitation. The characteristic
frequency of the plasmon has a narrower energy distribution in comparison
to the shake-up but both are similar in nature and significantly alter
the spectrum at low energies.  Further, we have shown that for small
systems the satellite region merges with the QP peak and shifts the
apparent photoemission peak maximum to lower energies. The QP
energies and photoemission maxima thus differ for the systems studied
by as much as 0.6~eV.

The position of the satellite region is dictated by the QP energies and the frequency
of the collective excitation. For small and low dimensional systems, surface and bulk plasmon resonances contribute to the satellites
(leading to inhomogeneous broadening of the satellites). With increasing size the higher energy bulk plasmon coupling dominates.  For small and intermediate systems, the maximum of the satellite decreases in energy  and is affected by the plasmon resonance energy. For big systems, the maximum shows a slight increase due to changes in the main part of the QP spectrum.

\acknowledgements This work was supported by the Center for Computational
Study of Excited-State Phenomena in Energy Materials (C2SEPEM) at the Lawrence
Berkeley National Laboratory, which is funded by the U.S. Department
of Energy, Office of Science, Basic Energy Sciences, Materials Sciences
and Engineering Division under Contract No. DE-AC02-05CH11231, as
part of the Computational Materials Sciences Program. The calculations
were performed as part of the XSEDE computational project TG-CHE160092
\cite{towns2014xsede}. This research used resources of the National Energy Research Scientific Computing Center, a DOE Office of Science User Facility supported by the Office of Science of the U.S.~Department of Energy under Contract No.~DE-AC02-05CH11231.

\bibliographystyle{unsrt}
\bibliography{library}

\end{document}